\documentclass[aps,prd,showpacs,twocolumn]{revtex4}
\usepackage{graphicx}

\begin{document}

\title{CPT Tests: Kaon vs Neutrinos\cite{acknowledgements}}

\author{Hitoshi Murayama}

\affiliation{Department of Physics,
     University of California, Berkeley, California 94720}
\affiliation{Theoretical Physics Group,
     Ernest Orlando Lawrence Berkeley National Laboratory\\
     University of California, Berkeley, California 94720}

\begin{abstract}
  CPT violation has an impressive limit in the neutral kaon system
  $|m(K^0) - m(\overline{K}^0)| < 10^{-18} m_K = 0.50 \times
  10^{-18}$~GeV.  However, if viewed as a constraint on the
  mass-squared, the bound appears weak, $|m^2(K^0) -
  m^2(\overline{K}^0)| < 0.25$~eV$^2$.  We point out that neutrino
  oscillation offers better limits on CPT violation in this case.  The
  comparison of solar and rector neutrino results puts the best limit
  on CPT violation by far, $|\Delta m^2_\nu - \Delta m^2_{\bar{\nu}}|
  < 1.3 \times 10^{-3}$~eV$^2$ (90\% CL).
\end{abstract}
\pacs{~}
\maketitle

The CPT theorem is one of the few solid predictions of the
relativistic local quantum field theory \cite{CPT}.  In particular, it
states that a particle and its anti-particle must have the same mass
and lifetime.  It is based on three reasonable assumptions:
\begin{itemize}
\item Lorentz invariance,
\item Hermiticity of the Hamiltonian,
\item Locality.
\end{itemize}
If CPT is found violated, the implication to the fundamental physics
is enormous, as at least one of the three assumptions above must be
violated.  One way to prove the CPT theorem is by defining $S$-matrix
elements by analytic continuation of the Euclidean correlation
functions.  The CPT transformation is then achieved by the Euclidean
rotation that changes the sign of the (imaginary) time and all spatial
coordinates, and hence is a symmetry of the $S$-matrix elements.
String theory is normally argued to be CPT-conserving, as its
$S$-matrix elements are defined precisely in this fashion.  However,
it does not exclude the possibility of a spontaneous violation of the
CPT symmetry depending on the details of the low-energy limit.
Kostelecky and collaborators have a series of papers on possible CPT
violation based on this point of view \cite{Kostelecky}.  It was also
argued that it may be possible to break CPT in a field theory
by giving up locality but not the other two \cite{BL} (see, however,
Ref.~\cite{Greenberg} for criticisms).

Recently, a possible CPT violation which allows different masses for
particles and anti-particles has attracted attention in neutrino
oscillation phenomenology.  If three indications for the neutrino
oscillation, solar
\cite{Homestake,SAGE,GALLEX,SuperKsolar,SNONC,SNODN}, atmospheric
\cite{SuperKatmos}, and LSND \cite{LSND}, are all correct, we
have to accommodate three mass-squared differences of quite different
orders of magnitudes, which is not possible within the three
generations of neutrinos.  A fourth kind of neutrino is usually
invoked to explain the data.  It has to be ``sterile'' so that it does
not violate the data from $Z^0$ decay at Large Electron Positron
collider (LEP): $N_\nu = 2.994 \pm 0.012$ \cite{PDG}.  However, recent
data from SNO requires $\nu_e$ oscillation into an active
(non-sterile) neutrino \cite{SNONC}, while the SuperKamiokande prefers
$\nu_\mu$ oscillation into $\nu_\tau$ \cite{tau}, leaving little room
for a sterile neutrino.  Further combined with older data from CDHSW
\cite{CDHSW}, Bugey \cite{Bugey}, even the extension with a sterile
neutrino does not help explain the data very much \cite{sterile}.
Yanagida and the author \cite{MY} have pointed out that we can explain
all data consistently allowing different mass spectra for neutrinos
and anti-neutrinos, because the solar neutrino oscillation is purely
in neutrinos while statistically significant evidence for oscillation
at LSND is in anti-neutrinos.  This observation was partially
motivated by the consistency between the LSND and SN1987A data.  This
possibility of CPT-violating neutrino mass spectra was elaborated
further by a series of works by Barenboim {\it et al}\/
\cite{Barenboim}.  (Indirect constraints are important only for the
Majorana case \cite{lepton}.)

Phenomenologically, a stringent limit exists on the CPT violation in
the neutral kaon system.  Thanks to the mixing between $K^0$ and
$\overline{K}^0$, the limit on the possible mass difference between
them is exceptionally strong \cite{PDG}:
\begin{equation}
  |m(K^0) - m(\overline{K}^0)| < 10^{-18} m_K 
  = 0.50 \times  10^{-9}~\mbox{eV}. \label{eq:mK}
\end{equation}
Given such a stringent limit, there does not appear much window for
CPT violation or improved tests \cite{BBM}.

We point out that the strength of the CPT limit from the neutral kaon
system may be misleading.  In lack of a concrete theory of CPT
violation, the limit Eq.~(\ref{eq:mK}) may be looked at as a limit on
the difference in {\it mass squared}\/ rather than the masses.  In
fact, a local Lagrangian field theory always has mass squared as a
natural parameter for bosons.  Also in relativistic kinematics, mass
squared is the natural parameter in Einstein's relation $E^2 =
\vec{p}^2 c^2 + m^2 c^4$ rather than the mass itself.  If
reinterpreted as a limit on the possible difference in mass squared,
it reads
\begin{equation}
  |m^2(K^0) - m^2(\overline{K}^0)| 
  < 0.25~\mbox{eV}^2. \label{eq:mK2}
\end{equation}
It is intriguing that the possible violation of CPT in quantum gravity
suppressed by the Planck scale may lead to an order of magnitude
$\langle v \rangle^2/M_{Pl} \sim 10^{-5}$~eV, which is well within the
above bound.

On the other hand, the neutrino oscillation experiments always measure
$\Delta m^2$, and cannot measure the masses themselves.  Yet, limits
on the difference $\delta \equiv \Delta m_\nu^2 - \Delta
m_{\bar{\nu}}^2$ can be obtained.  The SuperKamiokande collaboration
has studied the possible difference in neutrino and anti-neutrino
$\Delta m^2$ in atmospheric neutrino oscillations.  Their current
limit is \cite{Mauger}
\begin{equation}
  -7.5 \times 10^{-3}{\rm eV}^2 < \delta < 5.5 \times 10^{-3}{\rm eV}^2.
\end{equation}
This limit is much better than that from the kaon system.  We have to
note, however, that this limit assumes the same maximal mixing for
both neutrinos and anti-neutrinos.  The limit may be considerably
worse if this assumption is relaxed \cite{BL2}.

We find that the best limit comes from the comparison of the solar
neutrino data and the recent KamLAND result \cite{KamLAND}.  We
analyze the data within the two-flavor oscillation framework.
However, we emphasize that we cannot naively use the result of the
global fit to compare the preferred values of $\Delta m^2$ between
solar and reactor data.  It is because global fits are based on the
$\Delta \chi^2$ relative to the minimum and hence defines only the
relative probability, while throwing away information on which region
of the parameter space is excluded on the basis of the absolute
probability.  We have to find a way to obtain an absolute limit on the
parameter.

KamLAND has recently reported its initial result of a significant
deficit in the reactor anti-neutrino flux \cite{KamLAND}.  It
demonstrated a deficit in the reactor anti-neutrino flux, which we
interpret as neutrino oscillation.  Then we can speak of $\Delta
m^2_{\bar{\nu}}$.  Combined with the previous reactor experiments
CHOOZ \cite{CHOOZ} and Palo Verde \cite{PaloVerde}, we have a range of
$\Delta m_{\bar{\nu}}^2$ not excluded by the data:
\begin{equation}
  1.9 \times 10^{-5}{\rm eV}^2 < \Delta m_{\bar{\nu}}^2 
  < 1.1 \times 10^{-3}{\rm eV}^2
  \label{eq:reactor}
\end{equation}
at 90\%  CL {\it independent  of the mixing angle}\/  \cite{foot}.  We
emphasize that both  ends of the inequality are  the exclusion limits,
rather than the ``preferred range'' from the $\Delta \chi^2$ analysis.
Therefore,  this statement  has an  absolute meaning:  the probability
that  a value  of $\Delta  m_{\bar{\nu}}^2$ outside  this  range would
fluctuate and produce the observed data is less than 10\%.

As for the solar neutrino data, currently the Large Mixing Angle (LMA)
solution is the most preferred, while the LOW solution or Vacuum
oscillation (VAC) solution may exist at a higher confidence level.
From the analysis in \cite{BGP}, the goodness-of-fit is not
necessarily bad even for these solutions or Small Mixing Angle (SMA)
solution.  It is not clear we can set a lower limit on $\Delta
m_\nu^2$.  Fortunately for our purpose, it will suffice to have only
an upper bound on the $\Delta m_\nu^2$. 

SNO \cite{SNONC} convincingly established that the survival
probability of $^8$B neutrinos is about a third.  By naively combining
the reported numbers on solar neutrino fluxes with the charged-current
reaction $\phi_{CC}=1.76{}^{+0.06}_{-0.05}{}^{+0.09}_{-0.09}$ and with
the neutral-current reaction
$\phi_{NC}=5.09{}^{+0.44}_{-0.43}{}^{+0.46}_{-0.43}$, we find $P_{\rm
  surv} = \phi_{CC}/\phi_{NC} = 0.346 \pm 0.048$.  The upper bound at
90\% CL is $P_{\rm surv} < 0.425$.  It is important that it is less
than a half.  If the neutrinos oscillate purely in the vacuum, the
deficit would be at most a half in the case of the maximal mixing
\cite{vacuum}.  The deficit of two thirds is explained only by the
presence of the matter effect.  In order for the matter effect to be
important relative to the mass difference, $\Delta m_\nu^2$ is bounded
from above.  We would like to obtain a quantitative upper limit on
$\Delta m_\nu^2$ using this piece of information.

The Hamiltonian of the two-flavor neutrinos is
\begin{equation}
  H = \frac{\Delta m_\nu^2}{4p} \left(
    \begin{array}{cc}
      - \cos 2\theta & \sin 2\theta\\
      \sin 2\theta & \cos 2\theta
    \end{array} \right)
  + \left( 
    \begin{array}{cc}
      \sqrt{2} G_F n_e & 0\\
      0 & 0
    \end{array} \right).
\end{equation}
In this expression we dropped terms that are proportional to the
identity matrix as they are not important for the consideration of the
survival probabilities.  This Hamiltonian is time-dependent as the
electron number density $n_e$ changes in the course of neutrino
propagation.  The time evolution of the neutrino states is adiabatic
for high $\Delta m_\nu^2$, and hence we only need to study the
eigenstates of the Hamiltonian at the point of production ($n_e\approx
100 N_A/{\rm cm}^3$) and the detection ($n_e=0$) \footnote{The matter
  density of the Earth is negligible for high values of $\Delta
  m_\nu^2$ of our consideration here.}.  In the vacuum, the eigenstates
are given simply by
\begin{eqnarray}
  H \left( 
    \begin{array}{c}
      \cos \theta \\ -\sin\theta
    \end{array} \right) 
  &=& -\frac{\Delta m_\nu^2}{4p}
  \left(\begin{array}{c}
      \cos \theta \\ -\sin\theta
    \end{array} \right) , \\
  H \left( 
    \begin{array}{c}
      \sin \theta \\ \cos\theta
    \end{array} \right) 
  &=&  \frac{\Delta m_\nu^2}{4p}
  \left(\begin{array}{c}
      \sin \theta \\ \cos\theta
    \end{array} \right).
\end{eqnarray}
We choose the convention that $\Delta m_\nu^2 > 0$ without a loss of
generality, while the mixing angle is varied $0 < \theta < \pi/2$
\cite{Fogli,dark}.  On the other hand, in the core of the sun, 
\begin{eqnarray}
  H \left( 
    \begin{array}{c}
      \cos \theta_m \\ -\sin\theta_m
    \end{array} \right) 
  &=& E_-
      \left( 
        \begin{array}{c}
          \cos \theta_m \\ -\sin\theta_m
        \end{array} \right) , \nonumber \\
  H \left( 
    \begin{array}{c}
      \sin \theta_m \\ \cos\theta_m
    \end{array} \right) 
  &=& E_+
      \left( 
        \begin{array}{c}
          \sin \theta_m \\ \cos\theta_m
        \end{array} \right) ,
\end{eqnarray}
with
\begin{equation}
  E_\pm = \frac{\sqrt{2}G_F n_e}{2} \left(1 \pm {\sqrt{1 - 
            2 \Delta \cos 2\theta + 
            \Delta^2}}
      \right)
\end{equation}
and
\begin{equation}
  \Delta = \frac{\Delta m^2_\nu}{2p\sqrt{2}G_F n_e}.
\end{equation}
The mixing angle in the presence of matter is
\begin{equation}
  \tan \theta_m = \frac{1-\Delta \cos 2\theta 
    + \sqrt{1-2\Delta\cos 2\theta + \Delta^2}}{\Delta \sin 2\theta}.
\end{equation}
Because two different mass eigenstates decohere on the averaging over
the energy and the production region for this range of $\Delta
m_\nu^2$, we can obtain a very simple expression for the survival
probability
\begin{eqnarray}
  P_{\rm surv} &=& \cos^2 \theta \cos^2 \theta_m
  + \sin^2 \theta \sin^2 \theta_m \nonumber \\
  &=& \frac{1}{2} \left( 1 
    - \frac{\cos 2\theta (1-\Delta\cos 2\theta)}
    {\sqrt{1-2\Delta\cos 2\theta+\Delta^2}}\right).
  \label{eq:Psurv}
\end{eqnarray}

\begin{figure}[t]
  \centering \includegraphics[width=0.8\columnwidth]{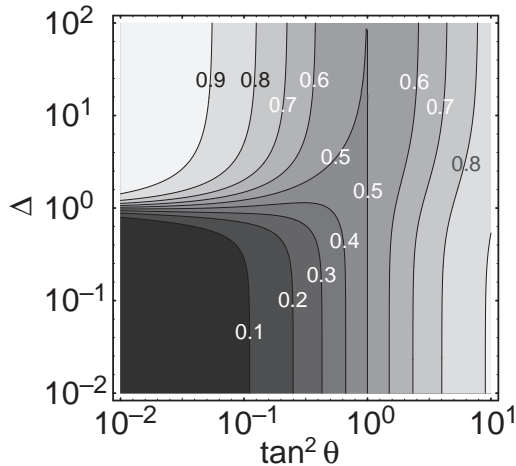}
  \caption{The contour plot of $P_{\rm surv}$ in Eq.~(\protect\ref{eq:Psurv}).}
  \label{fig:Psurv}
\end{figure}

The point here is that one cannot explain the reduction of the
electron neutrino flux to less than a half if $\Delta$ is too large.
One can show that the 90\% CL upper limit $P_{\rm surv}<0.425$
translates to $\Delta < 1.31$.  Therefore,
\begin{equation}
  \Delta m^2_\nu < 1.31 \times 2p\sqrt{2} G_F n_e.
\end{equation}
To be conservative, we use $n_e = 100 N_A{\rm cm}^{-3}$ at the core,
even though the production region of $^8$B neutrino is spread over
about a tenth of the solar radius.  We also conservatively take $p
\simeq 10$~MeV, the higher end of the $^8$B spectrum.  We then find
numerically,
\begin{equation}
  \Delta m^2_\nu < 2.0 \times 10^{-4}{\rm eV}^2.
  \label{eq:SNO}
\end{equation}

Now we combine Eqs.~(\ref{eq:reactor}) and (\ref{eq:SNO}) to obtain
a limit on possible CPT violation.  We, however, allow for the
possibility that the definition of $\Delta m^2$ may be different
between neutrinos and anti-neutrinos and hence they may have a
different sign.  Given this, the limit is
\begin{equation}
  |\delta| \equiv |\Delta m_\nu^2 - \Delta m_{\bar{\nu}}^2| 
  < 1.3 \times 10^{-3} {\rm  eV}^2 \qquad (90\% {\rm CL}). 
\end{equation}
Indeed this constraint is the world best bound on CPT violation in
mass-squared parameters so far.

The situation on the LSND evidence for neutrino oscillation remains
unresolved.  Naively, the consistency between solar neutrino data and
KamLAND seems to exclude the possibility of explaining LSND together
with other data using CPT violation within three generations.
However, the authors of Ref.~\cite{BL2} argued that the anti-neutrinos
are subdominant in atmospheric neutrino data and hence $\Delta m^2$ as
large as that of LSND is allowed for anti-neutrinos.  This point had
been criticized in \cite{GGMS}.  If LSND data stands, we may either
need more than one sterile neutrino \cite{SCS} or lepton number
violating muon decay \cite{BP}.  In the latter case, Mini-BooNE data
will not neither confirm or verify LSND data and the situation may
remain ambiguous.

In summary, we argued that the limit on CPT violation from the neutral
kaon system is not as strong as it appears once viewed as a constraint
on the mass-squared difference between kaon and anti-kaon.  Compared
to the kaon constraint, neutrino oscillation data provide much
stronger limits.  We derived a limit on $\delta = \Delta m^2_\nu -
\Delta m^2_{\bar{\nu}}$ quantitatively from SNO and KamLAND data, with
an emphasis on using the absolute probability rather than relying on
the $\Delta \chi^2$ analysis.  The obtained bound $|\delta| \equiv
|\Delta m_\nu^2 - \Delta m_{\bar{\nu}}^2| < 1.3 \times 10^{-3} {\rm
eV}^2$ (90\% CL) is currently the best limit on the possible CPT
violation in mass-squared of particles and anti-particles.


\begin{thebibliography}{99}
\bibitem{foot} The lower limit on $\Delta m^2$ from KamLAND
  \cite{KamLAND} is actually at 95\% CL, but this difference will not
  be important for later discussions.
\bibitem{vacuum}
The exception to this argument is when the energy dependence of the
survival probability is not averaged out, such as in the VAC
solution.  However, this case would require a much smaller $\Delta
m^2$ anyway and hence the upper bound we derive is still valid.  We
assume a large $\Delta m^2$ where energy dependence is averaged out
through the discussion.
\end{thebibliography}
\end{document}